\documentclass[showpacs,preprintnumbers,amsmath,amssymb,prl,superscriptaddress]{revtex4}	
\usepackage{graphicx} % Include figure files
\usepackage{dcolumn}  % Align table columns on decimal point
\usepackage{bm}       % bold math 
\usepackage{alltt}
\usepackage{calc,rotating,xspace}
\usepackage{pifont}

\newcommand{\ppbar}  {\ensuremath{p \bar{p}}}
\newcommand{\BsBsbar}{\ensuremath{B^0_s}-\ensuremath{\bar{B}^0_s}}
\newcommand{\Bs}     {\ensuremath{B_s^0}}
\newcommand{\Bsbar}  {\ensuremath{\bar{B}^0_s}} %IJK
\newcommand{\bs}     {\ensuremath{B_s}} %added by IJK

\newcommand{\Dsp}    {\ensuremath{D_s^+}}
\newcommand{\Bd}     {\ensuremath{B^0}}
\newcommand{\Bdbar}  {\ensuremath{\bar{B}^0}}

\newcommand{\keV}    {\ensuremath{\rm ke\kern -0.1em V}}
\newcommand{\MeV}    {\ensuremath{\rm Me\kern -0.1em V}}

\newcommand{\GeV}    {\ensuremath{\rm Ge\kern -0.1em V}}
\newcommand{\TeV}    {\ensuremath{\rm Te\kern -0.1em V}}

\newcommand{\ifb}    {\ensuremath{\mathrm{fb^{-1}}}}
\newcommand{\ips}    {\ensuremath{\mathrm{ps^{-1}}}}

\newcommand{\dms}    {\ensuremath{\Delta m_s}}
\newcommand{\dmd}    {\ensuremath{\Delta m_d}}
\newcommand{\Vtd}    {\ensuremath{V_{td}}}
\newcommand{\Vts}    {\ensuremath{V_{ts}}}

\newcommand{\VtdVts} {\ensuremath{|\Vtd/\Vts|}}
\newcommand{\Lrat}   {\ensuremath{\Lambda}}

\newcommand{\sye}[1]{\ensuremath{~\pm #1}}
\newcommand{\ase}[2]{\ensuremath{^{~+ #1}_{~- #2}}}
\newcommand{\massRatio} {\ensuremath{m_{\Bd}/m_{\Bs} = 0.98390}}
\newcommand{\deltaMdPdg}{\ensuremath{\dmd = 0.507\sye{0.005}\,\ips}}
\newcommand{\xiLat}     {\ensuremath{\xi = 1.21\ase{0.047}{0.035}}}
\newcommand{\deltaMsResultPRL}%
{\ensuremath{\dms =
  17.31\ase{0.33}{0.18}~({\rm stat})\sye{0.07}~({\rm syst})\,\ips}}
\newcommand{\deltaMsResult}%
{\ensuremath{\dms =
  17.77\sye{0.10}~({\rm stat})\sye{0.07}~({\rm syst})\,\ips}}
\newcommand{\VtdResult}%
{\ensuremath{\VtdVts =
  0.2060\sye{0.0007}~({\rm exp})\ase{0.0081}{0.0060}~({\rm theor})}}
\newcommand{\pvalue}{\ensuremath{8 \times 10^{-8}}}

\begin{document}
%%%%%%%%%%%%%%%%%%%%%%%%%%%%%%%%%%%%%%%%%%%%%%%%%%%%%%%%%%%%%%%%%%%%%%%%%%%%%%%

%% comment OUT next line for single-column-with-line-numbers
%%%{\hfill \today\ --- Version $0.0$}

%%%%%%%%%%%%%%%%%%%%%%%%%%%%%%%%%%%%%%%%%%%%%%%%%%%%%%%%%%%%%%%%%%%%%%%%%%%%%%%
% abstract and title
%%%%%%%%%%%%%%%%%%%%%%%%%%%%%%%%%%%%%%%%%%%%%%%%%%%%%%%%%%%%%%%%%%%%%%%%%%%%%%%

%%%%%1col\today\ {\bf 48 Hour Notice Version}

%\begin{LARGE}
%\begin{center}
%{\bf \boldmath Observation of {\BsBsbar} Oscillations from a Time-Dependent Measurement of {\dms}}
%\end{center}
%\end{LARGE}

%\title{\bf \boldmath Observation of {\BsBsbar} Oscillations
%from a Time-Dependent Measurement of {\dms}}
\title{\bf \boldmath Observation of {\BsBsbar} Oscillations}

%% UNcomment next line for single-column-with-line-numbers
%%%%%1col\maketitle

%%%author list goes here in two column version

\affiliation{Institute of Physics, Academia Sinica, Taipei, Taiwan 11529, Republic of China} 
\affiliation{Argonne National Laboratory, Argonne, Illinois 60439} 
\affiliation{Institut de Fisica d'Altes Energies, Universitat Autonoma de Barcelona, E-08193, Bellaterra (Barcelona), Spain} 
\affiliation{Baylor University, Waco, Texas  76798} 
\affiliation{Istituto Nazionale di Fisica Nucleare, University of Bologna, I-40127 Bologna, Italy} 
\affiliation{Brandeis University, Waltham, Massachusetts 02254} 
\affiliation{University of California, Davis, Davis, California  95616} 
\affiliation{University of California, Los Angeles, Los Angeles, California  90024} 
\affiliation{University of California, San Diego, La Jolla, California  92093} 
\affiliation{University of California, Santa Barbara, Santa Barbara, California 93106} 
\affiliation{Instituto de Fisica de Cantabria, CSIC-University of Cantabria, 39005 Santander, Spain} 
\affiliation{Carnegie Mellon University, Pittsburgh, PA  15213} 
\affiliation{Enrico Fermi Institute, University of Chicago, Chicago, Illinois 60637} 
\affiliation{Joint Institute for Nuclear Research, RU-141980 Dubna, Russia} 
\affiliation{Duke University, Durham, North Carolina  27708} 
\affiliation{Fermi National Accelerator Laboratory, Batavia, Illinois 60510} 
\affiliation{University of Florida, Gainesville, Florida  32611} 
\affiliation{Laboratori Nazionali di Frascati, Istituto Nazionale di Fisica Nucleare, I-00044 Frascati, Italy} 
\affiliation{University of Geneva, CH-1211 Geneva 4, Switzerland} 
\affiliation{Glasgow University, Glasgow G12 8QQ, United Kingdom} 
\affiliation{Harvard University, Cambridge, Massachusetts 02138} 
\affiliation{Division of High Energy Physics, Department of Physics, University of Helsinki and Helsinki Institute of Physics, FIN-00014, Helsinki, Finland} 
\affiliation{University of Illinois, Urbana, Illinois 61801} 
\affiliation{The Johns Hopkins University, Baltimore, Maryland 21218} 
\affiliation{Institut f\"{u}r Experimentelle Kernphysik, Universit\"{a}t Karlsruhe, 76128 Karlsruhe, Germany} 
\affiliation{High Energy Accelerator Research Organization (KEK), Tsukuba, Ibaraki 305, Japan} 
\affiliation{Center for High Energy Physics: Kyungpook National University, Taegu 702-701, Korea; Seoul National University, Seoul 151-742, Korea; and SungKyunKwan University, Suwon 440-746, Korea} 
\affiliation{Ernest Orlando Lawrence Berkeley National Laboratory, Berkeley, California 94720} 
\affiliation{University of Liverpool, Liverpool L69 7ZE, United Kingdom} 
\affiliation{University College London, London WC1E 6BT, United Kingdom} 
\affiliation{Centro de Investigaciones Energeticas Medioambientales y Tecnologicas, E-28040 Madrid, Spain} 
\affiliation{Massachusetts Institute of Technology, Cambridge, Massachusetts  02139} 
\affiliation{Institute of Particle Physics: McGill University, Montr\'{e}al, Canada H3A~2T8; and University of Toronto, Toronto, Canada M5S~1A7} 
\affiliation{University of Michigan, Ann Arbor, Michigan 48109} 
\affiliation{Michigan State University, East Lansing, Michigan  48824} 
\affiliation{Institution for Theoretical and Experimental Physics, ITEP, Moscow 117259, Russia} 
\affiliation{University of New Mexico, Albuquerque, New Mexico 87131} 
\affiliation{Northwestern University, Evanston, Illinois  60208} 
\affiliation{The Ohio State University, Columbus, Ohio  43210} 
\affiliation{Okayama University, Okayama 700-8530, Japan} 
\affiliation{Osaka City University, Osaka 588, Japan} 
\affiliation{University of Oxford, Oxford OX1 3RH, United Kingdom} 
\affiliation{University of Padova, Istituto Nazionale di Fisica Nucleare, Sezione di Padova-Trento, I-35131 Padova, Italy} 
\affiliation{LPNHE, Universite Pierre et Marie Curie/IN2P3-CNRS, UMR7585, Paris, F-75252 France} 
\affiliation{University of Pennsylvania, Philadelphia, Pennsylvania 19104} 
\affiliation{Istituto Nazionale di Fisica Nucleare Pisa, Universities of Pisa, Siena and Scuola Normale Superiore, I-56127 Pisa, Italy} 
\affiliation{University of Pittsburgh, Pittsburgh, Pennsylvania 15260} 
\affiliation{Purdue University, West Lafayette, Indiana 47907} 
\affiliation{University of Rochester, Rochester, New York 14627} 
\affiliation{The Rockefeller University, New York, New York 10021} 
\affiliation{Istituto Nazionale di Fisica Nucleare, Sezione di Roma 1, University of Rome ``La Sapienza," I-00185 Roma, Italy} 
\affiliation{Rutgers University, Piscataway, New Jersey 08855} 
\affiliation{Texas A\&M University, College Station, Texas 77843} 
\affiliation{Istituto Nazionale di Fisica Nucleare, University of Trieste/\ Udine, Italy} 
\affiliation{University of Tsukuba, Tsukuba, Ibaraki 305, Japan} 
\affiliation{Tufts University, Medford, Massachusetts 02155} 
\affiliation{Waseda University, Tokyo 169, Japan} 
\affiliation{Wayne State University, Detroit, Michigan  48201} 
\affiliation{University of Wisconsin, Madison, Wisconsin 53706} 
\affiliation{Yale University, New Haven, Connecticut 06520} 
\author{A.~Abulencia}
\affiliation{University of Illinois, Urbana, Illinois 61801}
\author{J.~Adelman}
\affiliation{Enrico Fermi Institute, University of Chicago, Chicago, Illinois 60637}
\author{T.~Affolder}
\affiliation{University of California, Santa Barbara, Santa Barbara, California 93106}
\author{T.~Akimoto}
\affiliation{University of Tsukuba, Tsukuba, Ibaraki 305, Japan}
\author{M.G.~Albrow}
\affiliation{Fermi National Accelerator Laboratory, Batavia, Illinois 60510}
\author{D.~Ambrose}
\affiliation{Fermi National Accelerator Laboratory, Batavia, Illinois 60510}
\author{S.~Amerio}
\affiliation{University of Padova, Istituto Nazionale di Fisica Nucleare, Sezione di Padova-Trento, I-35131 Padova, Italy}
\author{D.~Amidei}
\affiliation{University of Michigan, Ann Arbor, Michigan 48109}
\author{A.~Anastassov}
\affiliation{Rutgers University, Piscataway, New Jersey 08855}
\author{K.~Anikeev}
\affiliation{Fermi National Accelerator Laboratory, Batavia, Illinois 60510}
% ...
\collaboration{CDF Collaboration}
\noaffiliation

\begin{abstract}
 We report the observation of {\BsBsbar} oscillations from a
 time-dependent measurement of the {\BsBsbar} oscillation frequency {\dms}.
% We use 1~{\ifb} of data from $p\bar{p}$ collisions at $\sqrt{s}=1.96~\TeV$
% collected with the CDF\,II detector at the Fermilab Tevatron.
% The sample contains signals of 5700 fully reconstructed hadronic {\bs}
% decays, 3150 partially reconstructed hadronic {\bs} decays, 
% and 61\,500 partially reconstructed semileptonic {\bs} decays.
 Using a data sample of 1~\ifb\ of $p\bar{p}$ collisions at
 $\sqrt{s}=1.96~\TeV$
 collected with the CDF\,II detector at the Fermilab Tevatron,
 we find signals of 5600 fully reconstructed hadronic {\bs} decays,
 3100 partially reconstructed hadronic {\bs} decays, 
 and 61\,500 partially reconstructed semileptonic {\bs} decays.
 We measure the probability as a function of proper decay time that the
 {\bs} decays with the same, or opposite, flavor as the flavor at
 production, and we find a signal for {\BsBsbar} oscillations.
 The probability that random fluctuations could produce a comparable signal 
 is {\pvalue}, which exceeds $5\sigma$ significance.
 We measure {\deltaMsResult} and extract {\VtdResult}.
\end{abstract}

%% add PACS numbers
%% comment out \pacs for one column version with line numbers
\pacs{14.40.Nd, 12.15.Ff, 12.15.Hh, 13.20.He}

%% comment OUT next line for single-column-with-line-numbers
\maketitle

%% UNcomment next line for single-column-with-line-numbers
%\clearpage

%%%%%%%%%%%%%%%%%%%%%%%%%%%%%%%%%%%%%%%%%%%%%%%%%%%%%%%%%%%%%%%%%%%%%%%%%%%%%%%%
% introduction
%%%%%%%%%%%%%%%%%%%%%%%%%%%%%%%%%%%%%%%%%%%%%%%%%%%%%%%%%%%%%%%%%%%%%%%%%%%%%%%%
 Since the first observation of particle-antiparticle transformations
 in neutral $B$~mesons in 1987~\cite{UA1_ARGUS},
 the determination of the {\BsBsbar} oscillation frequency {\dms}
 from a time-dependent measurement of {\BsBsbar} oscillations
 has been a major objective of experimental particle physics~\cite{MIXING}.
 This frequency can be used to extract the magnitude of {\Vts},
 one of the nine elements
 of the Cabibbo-Kobayashi-Maskawa (CKM) matrix~\cite{CKM}.
% These elements are fundamental parameters in the standard electroweak model.
% Neutral $B$ mesons ($b\bar{q}$, with $q=d,s$ for {\Bdbar},
% {\Bsbar}) oscillate from particle to antiparticle due to 
% flavor-changing weak interactions with a frequency proportional
% to the mass difference $\Delta m_q$  between the two mass eigenstates
% $B^0_{q,H}$ and $B^0_{q,L}$~\cite{MIXING}.
% These mass differences can be used to extract the elements
% {\Vtd} and {\Vts} of the Cabibbo-Kobayashi-Maskawa (CKM) matrix~\cite{CKM},
% which are fundamental parameters in the standard electroweak model.
% The mass difference $\Delta m_d$ is precisely determined:
% $\deltaMdPdg$~\cite{PDG2006}.
%%% Recently, we reported the first precise measurement of {\dms}:
%%% {\deltaMsResultPRL}~\cite{CDF-BSMIX-2006},
%%% based on the analysis of 1~\ifb of data collected with the CDF\,II detector
%%% at the Fermilab Tevatron. 
%%% The probability that random fluctuations would produce a comparable signal
%%% was 0.2\% ($3\sigma$), which is too large to claim an observation.
 Recently, we reported~\cite{CDF-BSMIX-2006} the strongest evidence to date
 of the direct observation of {\BsBsbar} oscillations.
 That analysis used 1~\ifb\ of data collected with the CDF\,II detector at
 the Fermilab Tevatron, and the probability that random fluctuations would
 produce a comparable signal was 0.2\%, corresponding to $3\sigma$ signal
 significance.
 This level of significance is insufficient to claim a firm observation,
 however, under the oscillation hypothesis we determined {\deltaMsResultPRL}.
% In this Letter we update this measurement using the same data set
% and improved analysis techniques resulting in the first direct observation of
% $\bs$ oscillations.
 In this Letter we report an update of this measurement that uses the
 same data set with an improved analysis and reduces this probability to
 {\pvalue} ($>5\sigma$), yielding the first definitive observation of
 time-dependent {\BsBsbar} oscillations.

 We improve the analysis in Ref.~\cite{CDF-BSMIX-2006} by increasing
 the {\bs} signal yield
 and improving the performance of the methods
 used to identify the flavor ($b$ or $\bar{b}$) of the {\bs} at production.
 The previous analysis used {\bs} decays in hadronic
 ($\Bsbar\to D^+_s\pi^-$, $D^+_s\pi^-\pi^+\pi^-$) and semileptonic
 ($\Bsbar\to D^{+(*)}_s\ell^-\bar{\nu}_\ell$, $\ell=e$ or $\mu$)
 decay modes~\cite{CHARGECONJUGATE}. 
% The analysis in Ref.~\cite{CDF-BSMIX-2006} used {\bs} decays in hadronic
% ($\Bsbar\to D^+_s\pi^-$, $D^+_s\pi^-\pi^+\pi^-$) and semileptonic
% ($\Bsbar\to D^{+(*)}_s\ell^-\bar{\nu}_\ell$, $\ell=e$ or $\mu$)
% decay modes using charged particles only~\cite{CHARGECONJUGATE}. 
 We used $D^+_s\rightarrow \phi\pi^+$, $\bar{K}^{*}(892)^{0} K^+$, and
 $\pi^+\pi^-\pi^+$, with $\phi\rightarrow K^+K^-$
 and $\bar{K}^{*0}\rightarrow K^-\pi^+$.
 Several improvements lead to increased signal yields.
% We have improved the signal yields in several ways.
 We use particle identification techniques to find kaons from $D_s$~meson
 decays, allowing us to relax kinematic selection requirements on the $D_s$
 decay products.
 This results in increased efficiency for reconstructing the $D_s$
 while maintaining excellent signal to background.
 In the hadronic channels, we employ an artificial neural
 network (ANN) to improve candidate selection resulting in larger signal yields
 at similar or smaller background levels.
 The ANN selection makes it possible to use the additional decay sequence
 $\Bsbar\to D^+_s\pi^-\pi^+\pi^-$, with $D^+_s\to \pi^+\pi^-\pi^+$, as well. 
 We add significant statistics using partially reconstructed hadronic
 decays in which a photon or $\pi^0$ is missing:
 $\Bsbar\to D^{*+}_s\pi^-$, $ D^{*+}_s \to D^+_s \gamma/\pi^0$ and 
 $\Bsbar\to D^+_s\rho^-$, $\rho^- \to \pi^-\pi^0$,
 with $D^+_s\rightarrow \phi\pi^+$.
 Finally ANNs are used to enhance the performance of the methods
 used to identify the flavor of the {\bs} at production.
 With all these improvements, the effective statistical size of our data
 sample is increased by a factor of 2.5.

%%%%%%%%%%%%%%%%%%%%%%%%%%%%%%%%%%%%%%%%%%%%%%%%%%%%%%%%%%%%%%%%%%%%%%%%%%%%%%%%
% detector
%%%%%%%%%%%%%%%%%%%%%%%%%%%%%%%%%%%%%%%%%%%%%%%%%%%%%%%%%%%%%%%%%%%%%%%%%%%%%%%%

 The CDF\,II detector~\cite{DETECTOR_REFERENCE} consists of a magnetic
 spectrometer surrounded by electromagnetic and hadronic calorimeters
 and muon detectors.  The key components for this measurement are listed below.
 Precision determination of the decay point is
 provided by a seven-layer double-sided silicon-strip detector
 and a single-sided layer of silicon mounted directly on
 the beampipe at an average radius of 1.5\,cm.  A 96-layer
 drift chamber is used for both precision tracking and $dE/dx$
 particle identification.  Time-of-flight (TOF) counters surrounding
 the drift chamber are used to identify low-momentum charged kaons.
 A three-level trigger system selects, in real time, 
 {\ppbar} collisions containing charm and bottom hadrons by
 exploiting the kinematics of production and decay, and the long
 lifetimes of $D$ and $B$~mesons.  A crucial component of the trigger
 system for this measurement is the Silicon Vertex Trigger,
 which makes it possible to collect our large sample of {\bs} mesons
 in the fully or partially reconstructed hadronic decay
 modes, giving CDF unique sensitivity to {\bs} oscillations.

%%%%%%%%%%%%%%%%%%%%%%%%%%%%%%%%%%%%%%%%%%%%%%%%%%%%%%%%%%%%%%%%%%%%%%%%%%%%%%%
% data sample
%%%%%%%%%%%%%%%%%%%%%%%%%%%%%%%%%%%%%%%%%%%%%%%%%%%%%%%%%%%%%%%%%%%%%%%%%%%%%%%

\begin{figure}[htb]
\begin{center}
\includegraphics[width=0.49\linewidth, height = 0.49\linewidth]{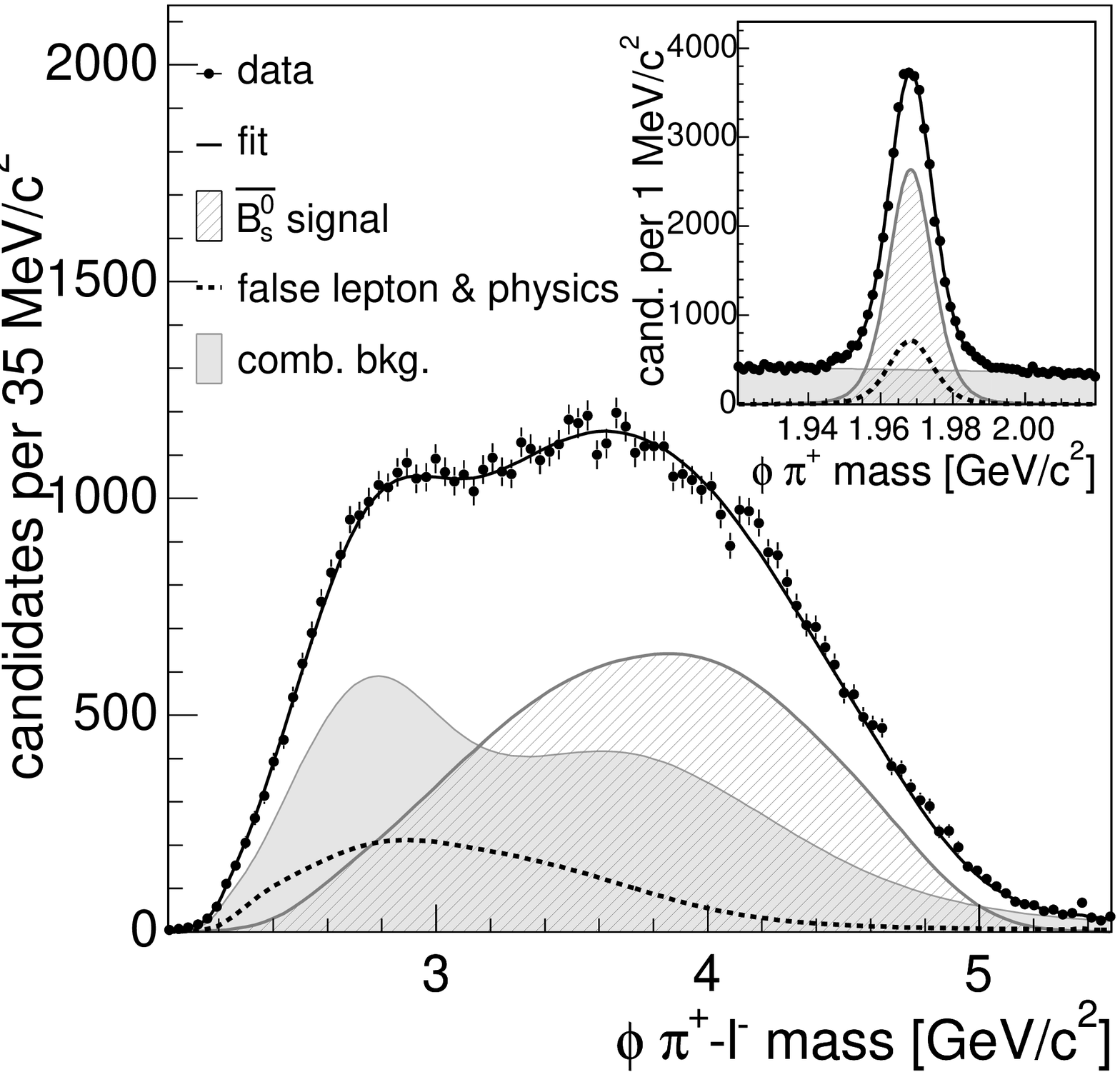}
\includegraphics[width=0.49\linewidth, height = 0.49\linewidth]{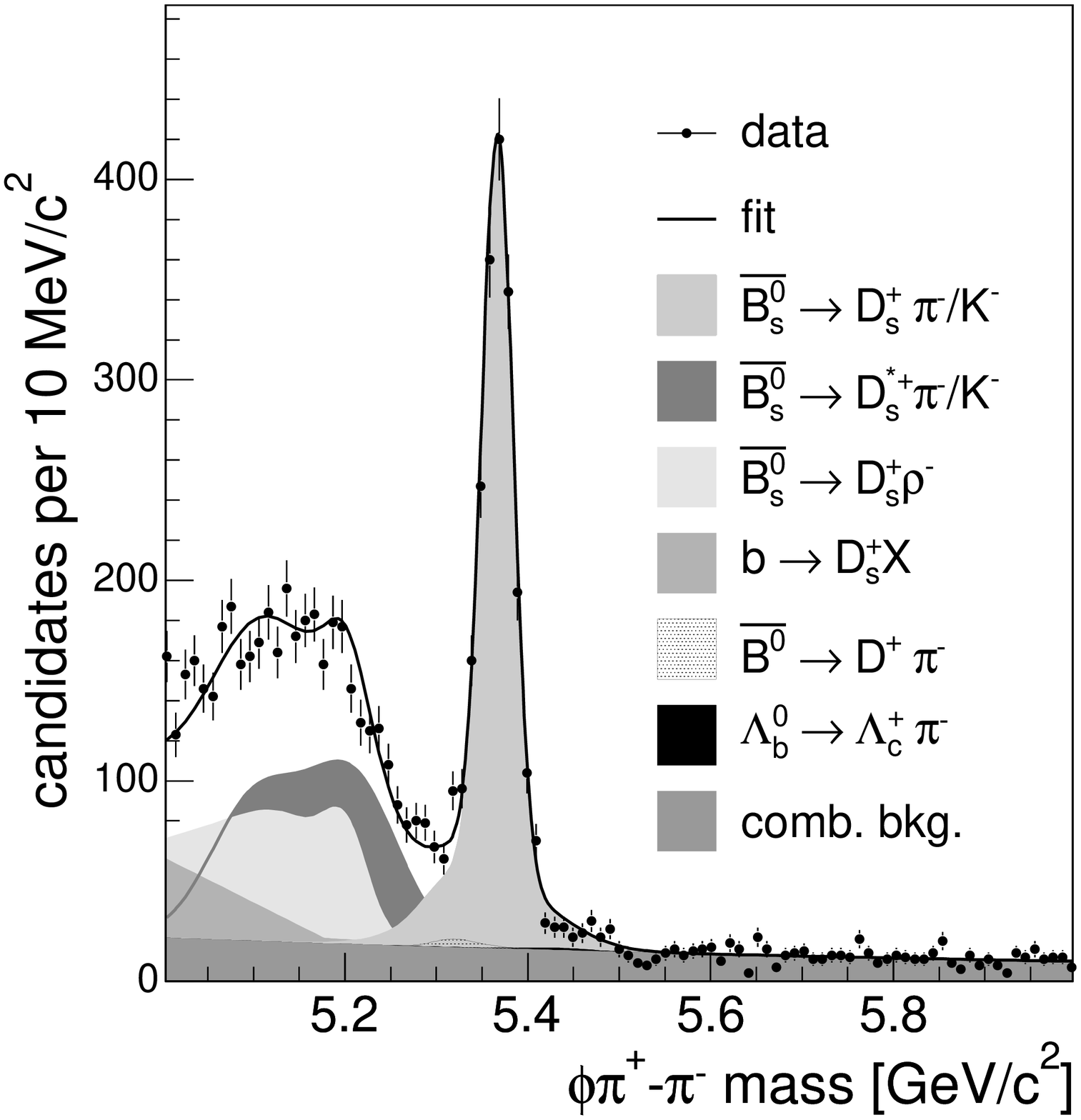}
\end{center}
\caption[] {
(Left panel) The invariant mass distributions for the $D^+_s(\phi\pi^+)$
candidates [inset] and the $\ell^-D^+_s(\phi\pi^+)$ pairs.
The contribution labelled ``false lepton \& physics'' refers to backgrounds
from hadrons mimicking the lepton signature combined with real $D_s$~mesons
and physics backgrounds such as
$B^0\rightarrow D_s^+ D^-$,
$D_s^+\rightarrow\phi\pi^+, D^-\rightarrow\ell^- X$.
(Right panel) The invariant mass distribution for
 $\Bsbar\to D^+_s(\phi \pi^+) \pi^-$ decays including the contributions from
 $\Bsbar \rightarrow D_s^{*+} \pi^-$ and $\Bsbar
  \rightarrow \Dsp \rho^-$.
 In this panel, signal contributions are drawn added on top of
 the combinatorial background.
}
\label{fig:invariantmass}
\end{figure}

To reconstruct {\Bsbar} candidates, we first select $D^+_s$ candidates.
These $D^+_s$ candidates are combined with one or three additional
charged particles to form
$D^+_s\ell^-$, $D^+_s\pi^-$, or $D^+_s\pi^-\pi^+\pi^-$ candidates.
In the previous analysis, we reduced combinatorial backgrounds by
applying requirements on selection quantities such as
the minimum $p_T$~\cite{DEFINE_PT} of the {\Bsbar} and its decay products,
and the quality of the reconstructed {\Bsbar} and $D^+_s$ decay points
and their displacement from the $p\bar{p}$ collision position.
In this analysis, we add a kaon identification likelihood
formed from TOF and $dE/dx$ information.
For decay modes with kaons in the final state, we use this likelihood to
reduce combinatorial background from random pions
or physics backgrounds such as $D^+\rightarrow K^-\pi^+\pi^+$.
In~\cite{CDF-BSMIX-2006}, we vetoed $D^+_s$ candidates consistent with the
$D^+$ mass hypothesis, which resulted in a substantial loss of signal
efficiency.
Kaon identification makes it possible to relax kinematic requirements
(charged particle $p_T$ and the $D^+$ veto)
leading to a substantial increase in signal efficiency.

In the semileptonic channel,
the main gain is in the
$D^+_s\ell^-$, $D^+_s\rightarrow \bar{K}^{*}(892)^{0} K^+$
sequence, where the signal is increased by a factor of 2.2.
An additional gain in signal by a factor of 1.3 with respect to our previous
analysis comes from adding data selected with different trigger requirements.
In total the signal of 37\,000 semileptonic {\bs} decays in
\cite{CDF-BSMIX-2006} is increased to 61\,500, and the signal to background
improves by a factor of two in the sequences with kaons in the final state.
The distributions of the invariant masses of the
$D_s^+(\phi\pi^+)\ell^-$ pairs $m_{D_s\ell}$ and
the $D_s^+(\phi\pi^+)$ candidates are shown in Fig.~\ref{fig:invariantmass}.
We use $m_{D_s\ell}$ to help distinguish signal, which occurs at higher
$m_{D_s\ell}$, from combinatorial and physics
({\it e.g.}, double-charm decays of $B$~mesons) backgrounds.

\begin{figure}[htb]
\begin{center}
\includegraphics[width=\linewidth]{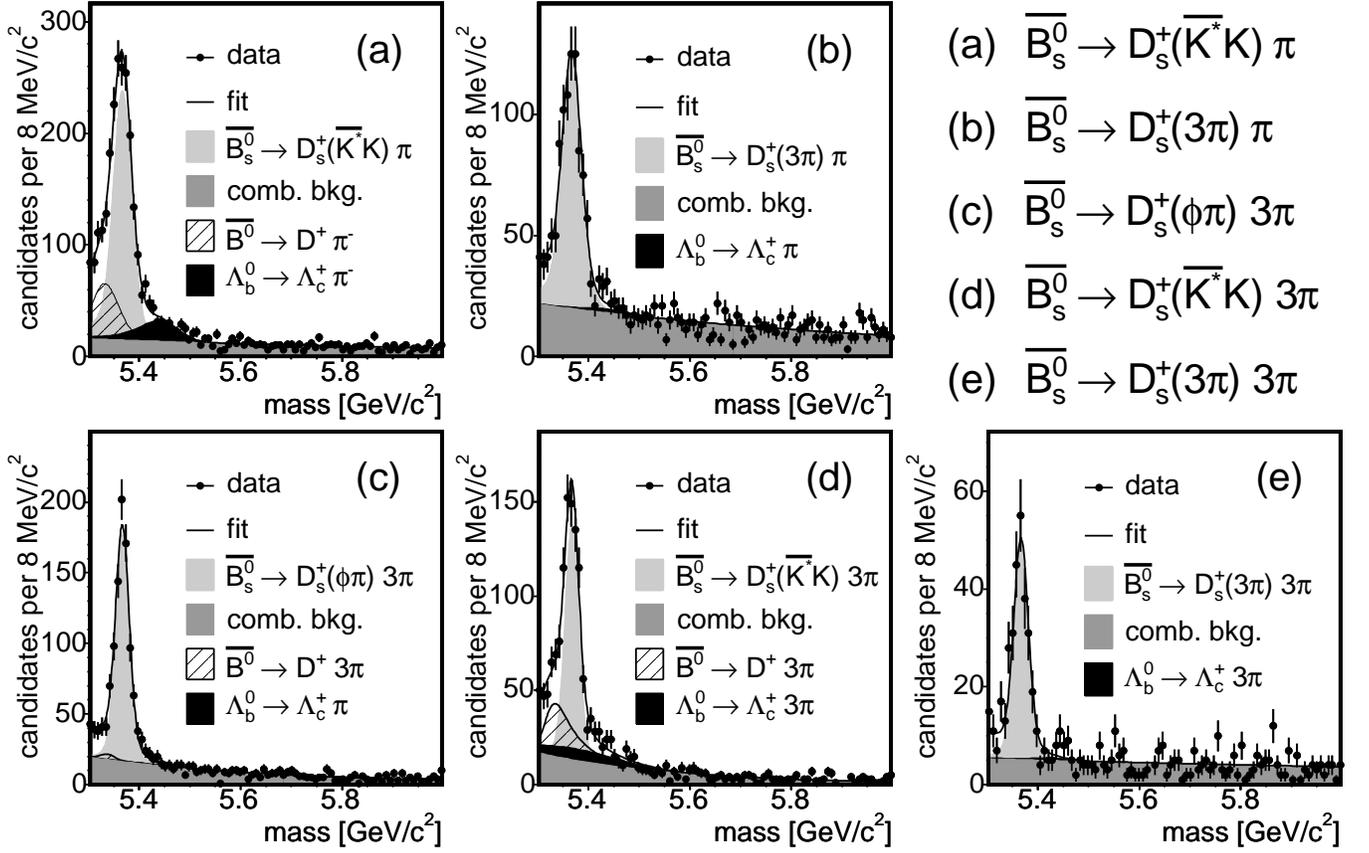}
\caption[] {
The invariant mass distributions for
$\Bsbar\to D^+_s\pi^-$ (top panels) and $D^+_s\pi^-\pi^+\pi^-$ (bottom panels).
Signal contributions are added on top of
the combinatorial background.
Contributions from partially reconstructed {\bs} decays are taken into account
in the fit and are not shown. 
}
\label{fig:invariantmass6}
\end{center}
\end{figure}

In the hadronic decay modes, we use an ANN to enhance
the signal selection of the previous analysis.
The ANN uses quantities such as the selection criteria listed above
as well as the kaon identification likelihood.
The network is trained using simulated signals generated with Monte Carlo
methods.
%that are passed through the full CDF detector simulation.
For combinatorial background, we use sideband regions in the mass
distribution of the {\bs} candidates from data.
In this analysis, we add the partially reconstructed signal
between 5.0 and 5.3\,GeV/$c^2$ from
$\Bsbar\to D^{*+}_s\pi^-$, $ D^{*+}_s \to D^+_s \gamma/\pi^0$
in which a photon or $\pi^0$ from the $D^{*+}_s$ is missing and 
$\Bsbar\to D^+_s\rho^-$, $\rho^- \to \pi^-\pi^0$ in which
a $\pi^0$ is missing.
The mass distributions for
$\Bsbar\to D^+_s\pi^-$, $D^+_s\rightarrow \phi\pi^+$ and the
partially reconstructed signals are shown in 
Fig.~\ref{fig:invariantmass}.
The mass distributions for the other five hadronic decay sequences
are shown in Fig.~\ref{fig:invariantmass6}.
In these modes, we require the masses of the candidates to be greater than
5.3\,GeV/$c^2$.
Candidates with masses greater than 5.5\,GeV/$c^2$ are used to construct
probability density functions (PDFs) for combinatorial background.
Table~\ref{table:signalyields} summarizes the signal yields.

\begin{table}
\begin{center}
\begin{tabular}{lccc}
\hline
\hline
Decay Sequence & Signal & S$/$B & gain w.r.t~\cite{CDF-BSMIX-2006} \\ \hline
$\Bsbar\to D^+_s(\phi\pi^+)\pi^-$
& 2000 & 11.3 &  13\% \\ %\hline
Partially reconstructed
& 3100 & 3.4 & n.a. \\ %\hline
$\Bsbar\to D^+_s(\bar{K}^*(892)^0 K^+)\pi^-$
& 1400 & 2.0 &  35\% \\ %\hline
$\Bsbar\to D^+_s(\pi^+\pi^-\pi^+)\pi^-$
&  700 & 2.1 &  22\% \\ %\hline
$\Bsbar\to D^+_s(\phi\pi^+)\pi^-\pi^+\pi^-$
&  700 & 2.7 &  92\% \\ %\hline
$\Bsbar\to D^+_s(\bar{K}^*(892)^0 K^+)\pi^-\pi^+\pi^-$
&  600 & 1.1 & 110\% \\ %\hline
$\Bsbar\to D^+_s(\pi^+\pi^-\pi^+)\pi^-\pi^+\pi^-$
&  200 & 2.6 & n.a.  \\ %\hline
%%%Satellites & 3300 & 3.4 & n.a. \\ %\hline
\hline
\hline
\end{tabular}
\caption[] {
Signal yields (S) and signal to background ratio (S/B) in the various
hadronic decay sequences. The gain refers to the percentage increase in
$S/\sqrt{S+B}$ relative to~\cite{CDF-BSMIX-2006}.}
\label{table:signalyields}
\end{center}
\end{table}

%%%%%%%%%%%%%%%%%%%%%%%%%%%%%%%%%%%%%%%%%%%%%%%%%%%%%%%%%%%%%%%%%%%%%%%%%%%%%%%
% decay time
%%%%%%%%%%%%%%%%%%%%%%%%%%%%%%%%%%%%%%%%%%%%%%%%%%%%%%%%%%%%%%%%%%%%%%%%%%%%%%%

 The reconstructed decay time in the $\bs$ rest frame is
 $t = m_{\bs}L_{T}/p_T^{\rm recon}$,
 where $L_{T}$ is the displacement of the $\bs$ decay point with respect to
 the primary vertex projected onto the $\bs$ transverse momentum vector, and
 $p_T^{\rm recon}$ is the transverse momentum
 of the reconstructed decay products.
 In the semileptonic and partially reconstructed hadronic decays, we correct
 $t$ by a factor $\kappa = p_T^{\rm recon}/p_T(\bs)$
 determined with Monte Carlo simulation (Fig.~\ref{fig:kappa-ctres}).
% This factor is the fraction of the total {\bs} $p_T$
% carried by the reconstructed decay products.
% To improve the decay-time resolution, we use event-by-event
% primary vertex position measurements when computing the $\bs$
% vertex displacement.
% The signal decay-time distribution is modeled with
% $P(t_i,\sigma_{t_i})
% =  \varepsilon(t_i) \int \Gamma_s e^{-\Gamma_s t'} 
%    {\cal G}(t' - t_i, \sigma_{t_i}) dt'$,
% where $t_i$ is the measured decay
% time of the $i$th candidate, $\Gamma_s$ is the $\bs$ decay width,
% ${\cal G}(x-\mu,\sigma)$ is a Gaussian distribution of the random variable
% $x$ with mean $\mu$ and width $\sigma$, and $\sigma_{t_i}$
% is the estimated candidate decay-time resolution.
% The decay-time efficiency function $\varepsilon(t)$
% describes trigger and selection biases on the decay-time distribution
% and is determined from Monte Carlo simulation.
% For semileptonic decays, the $\kappa$ distribution is determined from
% Monte Carlo simulation
% and is convoluted with the signal decay-time distribution.
% The missing transverse momentum from unreconstructed particles in the
% semileptonic decays is an important contribution to the decay-time resolution.
% To reduce this contribution and make optimal use of the semileptonic decays,
% we determine the $\kappa$ distribution
% as a function of $m_{D_s\ell}$.
% The r.m.s.~width of the $\kappa$ distribution is 3\% (20\%) for
% $m_{D_s\ell} = 5.2~\GeVcc$ ($3.0~\GeVcc$)(see Fig.~\ref{fig:kappa-ctres}).

 The decay time resolution $\sigma_t$ has contributions from the momentum of
 missing decay products (due to the spread of the distribution of $\kappa$)
 and from the uncertainty on $L_T$.
 The uncertainty due to the missing momentum increases with proper decay time
 and is an important contribution to $\sigma_t$ in the semileptonic decays.
% in the semileptonic decays, this uncertainty is an important contribution
% to $\sigma_t$.
 To reduce this contribution and make optimal use of the semileptonic decays,
 we determine the $\kappa$ distribution as a function of $m_{D_s\ell}$
 (Fig.~\ref{fig:kappa-ctres}).
 We estimate the contribution from the uncertainty on $L_T$ to $\sigma_t$ for
 each candidate using the measured track parameters and their estimated
 uncertainties.
% We calibrate this estimate using a large sample of prompt
% $D^+$~mesons, which we combine with one or three
% charged particles from the primary vertex to mimic signal topologies.

 The distribution of $\sigma_t$ for fully reconstructed
 decays has an average value of 87\,fs, which corresponds to one
 fourth of an oscillation period at $\dms = 17.8\,{\ips}$.
 The distribution is nearly Gaussian with an rms width of $31\,{\rm fs}$.
 For the partially reconstructed hadronic decays, the average $\sigma_t$
 is 97\,fs, and the addition to $\sigma_t$ due
 to the missing photon or $\pi^0$ is very small (Fig.~\ref{fig:kappa-ctres}).
 For semileptonic decays, $\sigma_t$ is worse due to decay
 topology and the much larger missing momentum of decay products that were not
 reconstructed.
 The increase of $\sigma_t$ with $t$ is illustrated
 in Fig.~\ref{fig:kappa-ctres} for different ranges of $m_{D_s\ell}$.  
% For example, at $t=0$, $\sigma_t=100$\,fs (200\,fs) for 
% $m_{D_s\ell} = 5.2~\GeVcc$ ($3.0~\GeVcc$) and increases to
% $\sigma_t=115$\,fs (380\,fs) at $t=1.5$\,ps(see Fig.~\ref{fig:ctres}).
% Similarly the decay-time
% resolution for partially reconstructed hadronic decays varies from
% 97 to 107\,fs.

%%%%%%%%%%%%%%%%%%%%%%%%%%%%%%%%%%%%%%%%%%%%%%%%%%%%%%%%%%%%%%%%%%%%%%%%%%%%%%%%
% flavor
%%%%%%%%%%%%%%%%%%%%%%%%%%%%%%%%%%%%%%%%%%%%%%%%%%%%%%%%%%%%%%%%%%%%%%%%%%%%%%%%
\vspace{0.5 cm}
\begin{figure}[htb]
\begin{center}
%%%%%note added for submitted version
%%%%%apparently cygwin does not distinguish kFactors.eps and KFactors.eps
%%%%%but linux does
%%%\includegraphics[width=0.49\linewidth, height = 0.49 \linewidth]{kFactors_prl.eps}
\includegraphics[width=0.49\linewidth, height = 0.49 \linewidth]{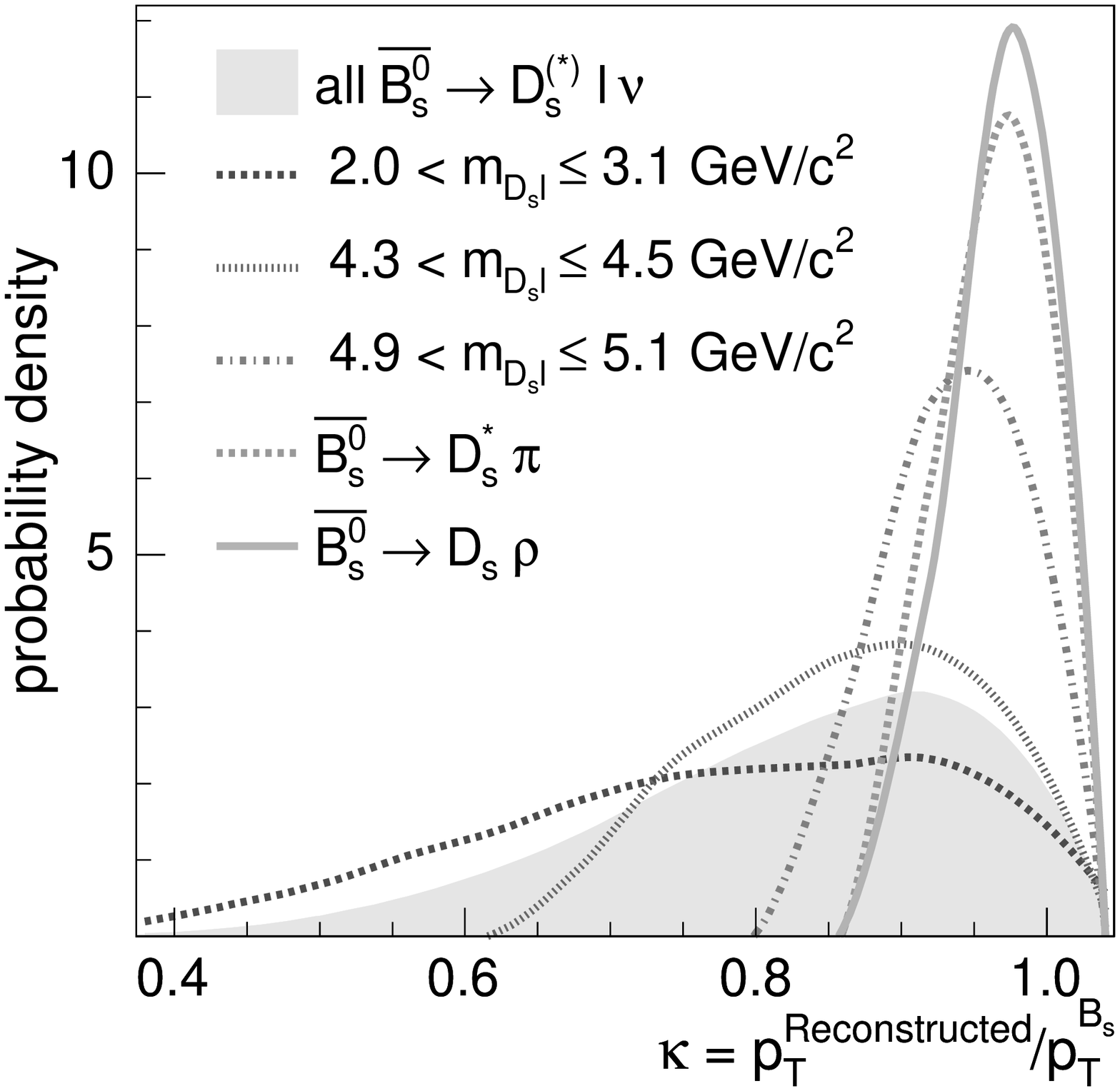}
\includegraphics[width=0.49\linewidth, height = 0.49 \linewidth]{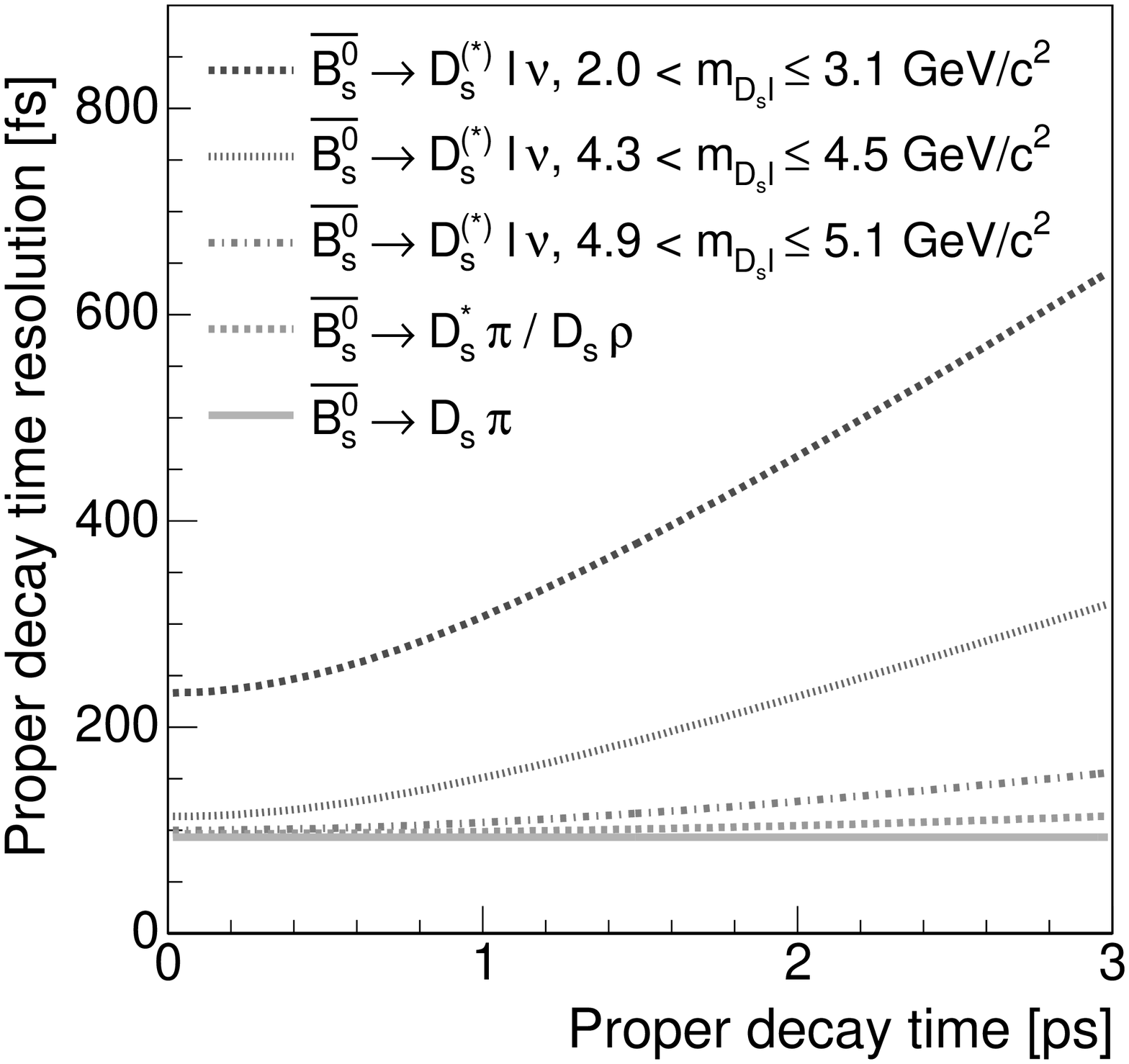}
\caption[] {
(Left panel) The distribution of the correction factor $\kappa$ in
semileptonic and partially reconstructed hadronic decays from Monte
Carlo simulation.  (Right panel) The average proper decay time
resolution for {\bs} decays as a function of proper
decay time.  }
\label{fig:kappa-ctres}
\end{center}
\end{figure}

 The flavor of the {\bs} at production is determined using both
 opposite-side and same-side flavor tagging techniques.
 The effectiveness $Q\equiv\epsilon {\cal D}^2$ of these techniques
 is quantified with an efficiency $\epsilon$, the fraction of signal candidates
 with a flavor tag, and a dilution ${\cal D}\equiv 1-2w$,
 where $w$ is the probability that the tag is incorrect.

 At the Tevatron, the dominant $b$-quark production mechanisms produce
 $b\bar{b}$~pairs.
 Opposite-side tags infer the production flavor of the {\bs} from the
 decay products of the $b$~hadron produced from the other $b$~quark
 in the event. 
 In the previous analysis, we used lepton ($e$ and $\mu$) charge
 and jet charge as tags, and if both types of tag were present,
 we used the lepton tag.
% building on
% techniques developed for a CDF Run\,I measurement of
% {\dmd}~\cite{OWEN_PRD}.
 In this improved analysis, we add an opposite-side
 flavor tag based on the charge
 of identified kaons, and we combine the information from the kaon, lepton,
 and jet charge tags using an ANN.
%to take advantage of correlations between
% tag decisions for candidates with more than one opposite-side flavor tag.
 % The dilution of opposite-side flavor tags is expected to be independent of
% the type of $B$~meson that produces the hadronic or semileptonic decay.
%%%for submission, add reference to J. Piedra thesis 
 The dilution is measured in data~\cite{PIEDRA_THESIS}
 using large samples of $B^-$, which do not change flavor, and $\bar{B}^0$,
 which can be used after accounting for their well-known oscillation frequency.
 The combined opposite-side tag effectiveness improves by 20\% to
 $Q = 1.8\pm 0.1\,\%$.
%where the uncertainty is dominated by the statistics of the control samples.
 Most of the improvement is for candidates with both a lepton and
 jet-charge tag.

 Same-side flavor tags are based on the charges
 of associated particles produced in the fragmentation of the
 $b$~quark that produces the reconstructed {\bs}.
% In the simplest picture of fragmentation,
% a $\pi^+$ ($\pi^-$) accompanies the formation of a $B^-$ ($B^+$),
% a $\pi^-$ ($\pi^+$) accompanies a $\overline{B}^0$ ($B^0$),
% and a $K^-$ ($K^+$) accompanies a {\Bsbar} ({\Bs}).
% In Run\,I, CDF established this method of production flavor identification
% in measurements of {\dmd}~\cite{PETAR} and the $\mathit{CP}$
% symmetry violating parameter $\sin (2\beta)$~\cite{CDF_SIN2BETA}.
 In the previous analysis, we used a same-side tag based on our
 kaon particle-identification likelihood; here
 we use an ANN to combine our kaon particle-identification likelihood with
 kinematic quantities of the kaon candidate into a single tagging variable $T$.
 Tracks close in phase space to the $\bs$ candidate are
 considered as same-side kaon tag candidates, and the track with the
 largest value of $T$ is selected as the tagging track.
% The performance of the same-side kaon tag for {\Bsbar} is expected to be
% different than for $B^-$ and $\overline{B}^0$.
 We predict the dilution of the same-side tag using simulated data samples
 generated with the {\sc pythia} Monte Carlo~\cite{PYTHIA} program.
 The predicted fractional gain in $Q$ from using the ANN is 10\%.
 Control samples of $B^-$ and $\bar{B}^0$ are used to validate
 the predictions of the simulation. 
 The effectiveness of this flavor tag increases with the $p_T$ of the
 {\Bsbar}; we find $Q = 3.7\%$ ($4.8\%$) in the hadronic (semileptonic)
 decay sample.
 The fractional uncertainty on $Q$ is approximately 25\%~\cite{CDF-BSMIX-2006}.
% This uncertainty is dominated by the differences between data and
% simulation for kaons found close in phase space to the {\Bsbar}~\cite{DENYS}
% and for the performance of the same-side kaon tag when applied to $B^-$.
 If both a same-side tag and an opposite-side tag are present,
 we combine the information from both tags assuming they are independent.
% The addition of the same-side kaon tag increases the effective sample
% statistics by more than a factor of three.
 
%%%%%%%%%%%%%%%%%%%%%%%%%%%%%%%%%%%%%%%%%%%%%%%%%%%%%%%%%%%%%%%%%%%%%%%%%%%%%%%
% fit and results
%%%%%%%%%%%%%%%%%%%%%%%%%%%%%%%%%%%%%%%%%%%%%%%%%%%%%%%%%%%%%%%%%%%%%%%%%%%%%%%

 We use an unbinned maximum likelihood fit to search for {\bs} oscillations. 
 The likelihood combines mass, decay time, decay-time resolution, and
 flavor tagging information for each candidate, and includes terms for
 signal and each type of background.  Details of the fit are described
%%%for arXiv, add Leonardo thesis here.
 in~\cite{CDF-BSMIX-2006,LEONARDO_THESIS}. 
% The fit is done in three stages. 
% First, a combined mass and decay-time fit is performed to separate
% signal from background and to fix mass and decay-time models.
% Combined fits for $\bs$ mass
% and decay width in hadronic samples
% and for decay width in the semileptonic samples
% yield measurements consistent with established values
% \cite{PDG2006}.
% Next, flavor asymmetries are measured for background components.
% The last step is a fit for {\BsBsbar} oscillations; the mass and decay-time
% models and background asymmetries are fixed from the previous
% two stages.

% The signal PDF has the general form:
% \begin{eqnarray*}
% &&{\cal S}_{\pm}(t_i,\sigma_{t_i}, {\cal D}_i) = \\
% &&\varepsilon(t_i)
% \int \frac{ \Gamma_s }{2} e^{-\Gamma_s t'}
% \left[ 1 \pm {\cal A} {\cal D}_i \cos (\Delta m_s t') \right ]
% {\cal G}( t_i - t', \sigma_{t_i} )\ dt',
% \end{eqnarray*}
% where ${\cal D}_i$ is the $i$th candidate
% dilution, and $t_i$, $\sigma_{t_i}$, $\cal G$, and $\varepsilon(t)$ have been
% defined previously. 
 Following the method described in \cite{MOSER}, we fit for the oscillation
 amplitude $\cal A$ while fixing $\Delta m_s$ to a probe value.
 The oscillation amplitude is expected to be consistent with
 ${\cal A} = 1$ when the probe value is the true oscillation frequency,
 and consistent with ${\cal A} = 0$ when the probe value is far from
 the true oscillation frequency.
 Figure~\ref{fig:amplitudeScan} shows the fitted value of the
 amplitude as a function of the oscillation frequency
 for the semileptonic candidates alone, the hadronic candidates alone,
 and the combination. 
% The sensitivity of the measurement is defined by the maximum value of
% {\dms } where ${\cal A}=1$ is excluded at 95\% C.L.~if
% the measured value of ${\cal A}$ were zero.
 The sensitivity~\cite{CDF-BSMIX-2006,MOSER} is 19.3\,{\ips} for the
 semileptonic decays alone, 30.7\,{\ips} for the hadronic decays alone,
 and 31.3\,{\ips } for all decays combined.
 At $\dms = 17.75~\ips $, the observed amplitude
 ${\cal A} = 1.21\pm 0.20~({\rm stat.})$
 is consistent with unity, indicating that the data
 are compatible with {\BsBsbar} oscillations with that frequency, while the
 amplitude is inconsistent with zero: ${\cal A}/\sigma_{\cal A}=6.05$,
 where $\sigma_{\cal A}$ is the statistical uncertainty on ${\cal A}$
 (the ratio has negligible systematic uncertainties).
% The negative amplitudes measured at frequencies slightly below and slightly
% above the peak frequency are expected and are due to the finite range in
% signal decay time that is imposed by the trigger and selection criteria.
% The systematic uncertainty on ${\cal A}$ is mainly due to uncertainties
% on $\sigma_{t_i}$ and ${\cal D}_i$.
% Since the effect of these uncertainties on ${\cal A}$ and $\sigma_{\cal A}$
% are correlated, the ratio ${\cal A}/\sigma_{\cal A}$ has negligible
% systematic uncertainty.
 The small uncertainty on ${\cal A}$ at $\dms = 17.75~\ips$
 is due to the superior decay-time resolution of the hadronic decay modes. 
% A separate amplitude scan for only the semileptonic decays is shown
% in figure~\ref{fig:amplitudeScan} which also shows a signal 
% at $\dms = 17.75~\ips $ consistent with the signal above.

\begin{figure}[ht]
\begin{center}
\includegraphics[width=\linewidth]{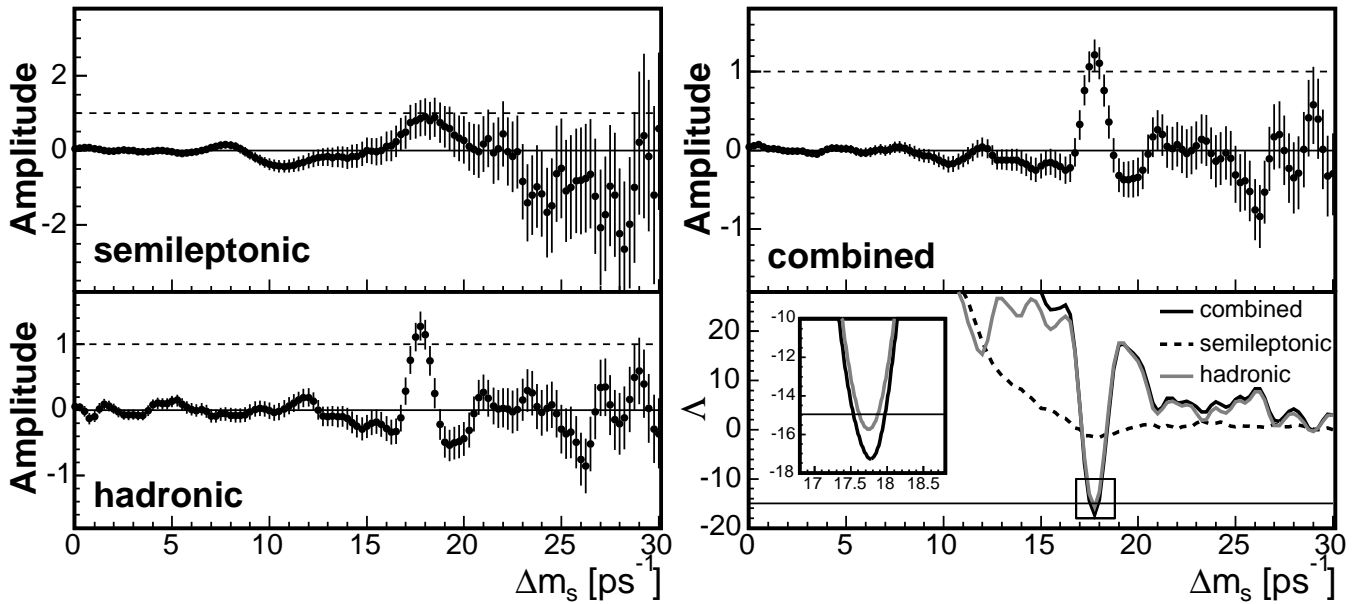}
\caption[] {
 The measured amplitude values and uncertainties versus the {\BsBsbar}
 oscillation frequency $\dms$. (Upper Left) Semileptonic decays only.
 (Lower Left) Hadronic decays only. (Upper Right) All decay modes combined. 
 (Lower Right)
 The logarithm of the ratio of likelihoods for amplitude equal to
 one and amplitude equal to zero,
 $\Lambda = \log [ {\cal L}^{{\cal A}=0} / {\cal L}^{{\cal A}=1}(\dms) ]$,
 versus the oscillation frequency.
 The horizontal line indicates the value $\Lambda = -15$ that corresponds
 to a probability of $5.7\times 10^{-7}$ (5$\sigma$)
 in the case of randomly tagged data. 
}
\label{fig:amplitudeScan}
\end{center}
\end{figure}

 We evaluate the significance of the signal using
 $\Lrat \equiv \log [ {\cal L}^{{\cal A}=0} / {\cal L}^{{\cal A}=1}(\dms)]$,
 which is the logarithm of the ratio of likelihoods for the hypothesis
 of oscillations (${\cal A}=1$) at the probe value and the hypothesis
 that ${\cal A}=0$, which is equivalent to random production flavor tags.
 Figure~\ref{fig:amplitudeScan} shows {\Lrat }
 as a function of {\dms }.
 Separate curves are shown for the semileptonic data alone (dashed),
 the hadronic data alone (light solid), and the combined data (dark solid).
 At the minimum $\dms=17.77~ \ips$, $\Lrat = -17.26$.
 The significance of the signal is the
 probability that randomly tagged data would produce a value 
 of {\Lrat } lower than $-17.26$ at any value of {\dms}.
 We repeat the likelihood scan 350 million times with random tagging decisions;
 28 of these scans have $\Lambda<-17.26$, corresponding to a probability of
 $\pvalue$ ($5.4\,\sigma$), well below $5.7 \times 10^{-7}$ ($5\,\sigma$).

\begin{figure}[h]
\begin{center}
\includegraphics[width=0.49\linewidth]{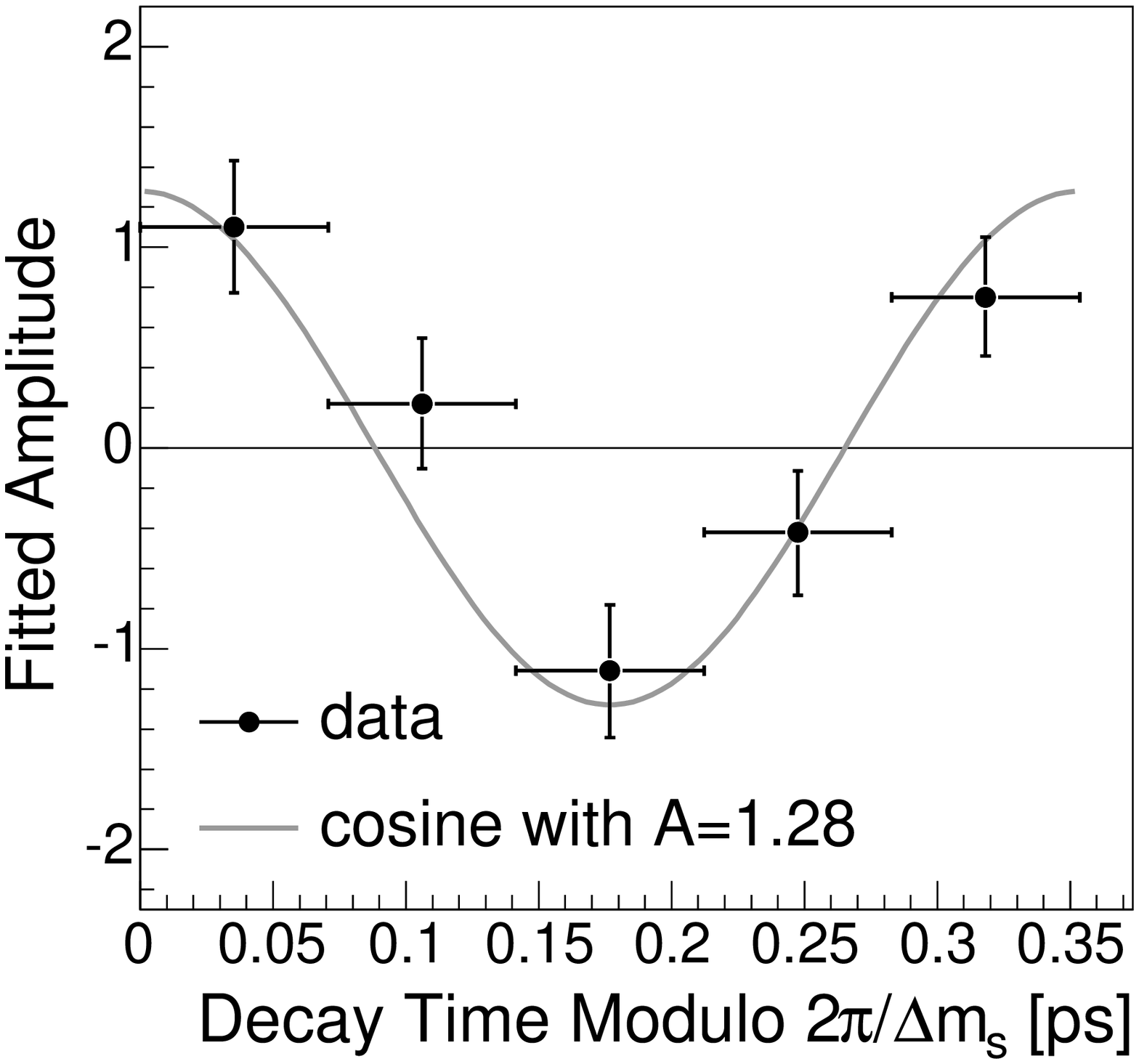}
\caption[] {
 The {\BsBsbar} oscillation signal measured in five bins of proper decay time
 modulo the measured oscillation period $2\pi/\dms$.
 The figure is described in the text.
}
\label{fig:asym}
\end{center}
\end{figure}

 %Given the observation of a {\BsBsbar} oscillation
 To measure {\dms}, we fix ${\cal A}=1$ and fit for the oscillation frequency.
 We find {\deltaMsResult}.
% All systematic uncertainties affecting ${\cal A}$
% are unimportant for {\dms}. 
 The only non-negligible systematic uncertainty on {\dms } is from
 the uncertainty on the absolute scale of the decay-time measurement.
 Contributions to this uncertainty include biases in the primary-vertex
 reconstruction due to the presence of the opposite-side $b$~hadron,
 uncertainties in the silicon-detector alignment, and biases in track fitting.
 The uncertainty on the correction $\kappa$ for the hadronic candidates with
 a missing photon or $\pi^0$ is included and has a negligible effect.

The {\BsBsbar} oscillations are depicted in Fig.~\ref{fig:asym}.
Candidates in the hadronic sample are collected in five bins
of proper decay time modulo the measured oscillation period $2\pi/\dms$.
In each bin, we fit for an amplitude (the points in Fig.~\ref{fig:asym})
using the likelihood function~\cite{CDF-BSMIX-2006},
which takes into account the effects of background,
flavor tag dilution and decay-time resolution for each candidate.
The curve shown in Fig.~\ref{fig:asym} is a cosine with an
amplitude of 1.28, which is the observed value in the
amplitude scan for the hadronic sample at $\dms=17.77$\,$\ips$.
As expected, the data are well represented by the curve.

 The measured {\BsBsbar} oscillation frequency is used to derive the ratio
 $\VtdVts  = 
\xi\sqrt{ \frac{\dmd}{\dms} \frac{m_{\Bs}}{m_{\Bd}} }$~\cite{PDG2006}.
 As inputs we use {\massRatio}~\cite{CDF_BS_MASS} with negligible
 uncertainty, {\deltaMdPdg}\,\cite{PDG2006} and
 {\xiLat}\,\cite{XI_LAT}. We find {\VtdResult}.

%%%%%%%%%%%%%%%%%%%%%%%%%%%%%%%%%%%%%%%%%%%%%%%%%%%%%%%%%%%%%%%%%%%%%%%%%%%%%%%
% conclusion
%%%%%%%%%%%%%%%%%%%%%%%%%%%%%%%%%%%%%%%%%%%%%%%%%%%%%%%%%%%%%%%%%%%%%%%%%%%%%%%
 
 %In conclusion, we present the first observation of {\BsBsbar} oscillations
 %from a decay-time dependent measurement of {\dms} and a 
 %precise value of $\dms$.
 %The value of $\dms$ is consistent with standard model expectations.
 %This result supersedes our previous measurement~\cite{CDF-BSMIX-2006}.

% In conclusion, we report the first observation of {\BsBsbar} oscillations
% with a significance that exceeds $5\sigma$ from a decay-time dependent
% measurement of {\dms} and a precise value of $\dms$.
% The value of $\dms$ is consistent with standard model expectations.
 In conclusion, we report the first observation of {\BsBsbar} oscillations
 from a decay-time dependent measurement of {\dms}.
 Our signal exceeds $5\sigma$ significance and yields a precise value of
 {\dms}, which is consistent with standard model expectations.
 This result supersedes our previous measurement~\cite{CDF-BSMIX-2006}.
 
% Our measured value of $\dms$ allows us to determine {\VtdVts} with
% and can be used to improve constraints on the
% unitarity of the CKM matrix and on scenarios involving new physics.

%%%% ACKNOWLEDGEMNTS %%%%%%%%%%%%%%%%%%%%%%%%%%%%%%%%%%%%%%%%%%%%%%%%%%%%%%%%%%%
We thank the Fermilab staff and the technical staffs of the participating
institutions for their vital contributions. This work was supported by the
U.S. Department of Energy and National Science Foundation; the Italian Istituto
Nazionale di Fisica Nucleare; the Ministry of Education, Culture, Sports,
Science and Technology of Japan; the Natural Sciences and Engineering Research
Council of Canada; the National Science Council of the Republic of China; the
Swiss National Science Foundation; the A.P. Sloan Foundation; the
Bundesministerium f\"ur Bildung und Forschung, Germany; the Korean Science and
Engineering Foundation and the Korean Research Foundation; the Particle Physics
and Astronomy Research Council and the Royal Society, UK;
the Institut National de Physique Nucleaire et Physique
des Particules/CNRS; the Russian Foundation
for Basic Research; the
Comisi\'on Interministerial de Ciencia y Tecnolog\'{\i}a, Spain; the
European Community's Human Potential Programme under contract
HPRN-CT-2002-00292; and the Academy of Finland.
%%%Obsolete version 
%We thank the Fermilab staff and the technical staffs of the participating
%institutions for their vital contributions. This work was supported by the
%U.S. Department of Energy and National Science Foundation; the Italian Istituto
%Nazionale di Fisica Nucleare; the Ministry of Education, Culture, Sports,
%Science and Technology of Japan; the Natural Sciences and Engineering Research
%Council of Canada; the National Science Council of the Republic of China; the
%Swiss National Science Foundation; the A.P. Sloan Foundation; the
%Bundesministerium f\"ur Bildung und Forschung, Germany; the Korean Science and
%Engineering Foundation and the Korean Research Foundation; the Particle Physics
%and Astronomy Research Council and the Royal Society, UK; the Russian Foundation
%for Basic Research; the Comisi\'on Interministerial de Ciencia y
%Tecnolog\'{\i}a, Spain; in part by the European Community's Human Potential
%Programme under contract HPRN-CT-2002-00292; and the Academy of Finland.

%%%%%%%%%%%%%%%%%%%%%%%%%%%%%%%%%%%%%%%%%%%%%%%%%%%%%%%%%%%%%%%%%%%%%%%%%%%%%%%
% references
%%%%%%%%%%%%%%%%%%%%%%%%%%%%%%%%%%%%%%%%%%%%%%%%%%%%%%%%%%%%%%%%%%%%%%%%%%%%%%%

%%%\input{bibliography.tex}

%%%%%%%%%%%%%%%%%%%%%%%%%%%%%%%%%%%%%%%%%%%%%%%%%%%%%%%%%%%%%%%%%%%%%%%%%%%%%%%

\begin{thebibliography}{99}
\bibitem{CHARGECONJUGATE}
%%
  The symbol {\bs} refers to the combination of
  {\Bsbar} and {\Bs} decays.
  References to a particular process imply that
  the charge conjugate process is included as well.
\bibitem{DEFINE_PT}
%%
  The transverse momentum $p_T$ is the magnitude of the component
  of the momentum perpendicular to the proton beam direction.
\end{thebibliography}
\end{document}
%%%%%%%%%%%%%%%%%%%%%%%%%%%%%%%%%%%%%%%%%%%%%%%%%%%%%%%%%%%%%%%%%%%%%%%%%%%%%%%